\documentstyle[12pt]{article}

\newcommand{\intg}{\int d^4x g^{\frac{1}{2}}}
\newcommand{\brs}{\hat{\delta}}
\newcommand{\dm}{D_{\mu}}
\newcommand{\dn}{D_{\nu}}
\newcommand{\mono}{F_{\mu \nu}^{a+}+q^{\dagger} \sigma_{\mu \nu} T^{a} q}
\newcommand{\D}{\dm \phi}
\newcommand{\qt}{q^{\dagger}}
\newcommand{\qtil}{\tilde{q}}

\newcommand{\la}{\lambda}
\newcommand{\x}{\chi}
\newcommand{\m}{\mu}
\newcommand{\n}{\nu}
\newcommand{\mn}{\m \n}
\newcommand{\Fp}{F^{+}}
\newcommand{\p}{\phi}
\newcommand{\pbar}{\bar{\phi}}
\newcommand{\al}{\alpha}
\newcommand{\ald}{\dot{\alpha}}
\newcommand{\psib}{\bar{\psi}}

\newcommand{\s}{\sigma}
\newcommand{\sbar}{\bar{\sigma}}
\newcommand{\be}{\begin{eqnarray}}
\newcommand{\ee}{\end{eqnarray}}
\newcommand{\gmn}{g^{\mu \nu}}
\newcommand{\ep}{\epsilon}
\newcommand{\epd}{\epsilon'}
\newcommand{\dse}{D^{2}_{\epsilon}}
\newcommand{\de}{D_{\mu \epsilon}}
\newcommand{\intD}{\int {\cal D}X}
\newcommand{\Dm}{\tilde{\cal D}_{\mu}}
\newcommand{\Ds}{\not \! \!{\tilde{\cal D}}}
\newcommand{\pmu}{\partial_{\mu}}
\newcommand{\pmuu}{\partial^{\mu}}

\newcommand{\DM}{{\cal D}_{\mu}}
\newcommand{\DN}{{\cal D}_{\nu}}
\newcommand{\DMS}{{\cal D}_{\mu}^{\ast}}
\newcommand{\DNS}{{\cal D}_{\nu}^{\ast}}
\newcommand{\cD}{{\cal D}}

\begin{document}
 


\begin{titlepage}
\vglue 3cm

\begin{center}
\vglue 0.5cm
{\Large\bf Reducible Connections 
\\in Massless Topological QCD and 4-manifolds} 
\vglue 1cm
{\large A.Sako} 
\vglue 0.5cm
{\it Department of Physics,Hokkaido University,Sapporo 060 ,Japan}
\baselineskip=12pt

\vglue 1cm
{\bf ABSTRACT}
\end{center}
{\rightskip=3pc
 \leftskip=3pc}

\noindent

A role of reducible connections in Non-Abelian Seiberg-Witten
 invariants is analyzed with 
massless Topological QCD where 
monopole is extended to non-Abelian groups version.  
By giving small external fields, we found that 
vacuum expectation value can be separated 
into a part from Donaldson theory
, a part from Abelian Monopole theory and a part from 
non-Abelian monopole theory.
As a by-product, 
we find identities of U(1) topological invariants.
In our proof, the duality relation and Higgs mechanism are not necessary.

\vspace{5cm}

\begin{flushleft}
\baselineskip=12pt
\hrule
\vspace{0.2cm}
{sako@particle.sci.hokudai.ac.jp}
\end{flushleft}

\end{titlepage}

\newpage

\section{Introduction}

Recently in differential topology of four manifolds there
have been new developements
by Seiberg-Witten monopole theory 
\cite{W-1}
  .
They conquered a difficulty of Donaldson
theory 
.
 Donaldson theory solved many problems of differential topology of 
4-manifolds about intersection form, polynomial invariants and so on
\cite{D-1}\cite{D-2}.
Donaldson theory is described by
 non-Abelian gauge theory, hence 
calculations are difficult.
Seiberg-Witten theory is easy for its computation
since it is Abelian gauge theory.
Donaldson invariants which is written by Kronheimer-Mrowka structure
formula are related with Seiberg-Witten invariants, and the relation has been
proved in several ways 
\cite{tyurin}\cite{ot}
 .
Hyun-J.Park-J.S.Park show the relation in path-integral formalism 
using massive Topological QCD \cite{HPP}.
Both Donaldson and Seiberg-Witten theory are 
understood as Topological field theory 
\cite{DW} \cite{TQCD}.
 Hyun-J.Park-J.S.Park computed path-integrals
 of massive Topological QCD and they found the way of separating it into 
two brunches that is Donaldson part and Seiberg-Witten part.
Symbolically the result of their computation is 
\be
\langle massive \;\;Topological \;\; QCD \rangle 
= \frac{a}{m^k} \left\{ \langle Donaldson \rangle
- \frac{b}{m^l}\langle Seiberg-Witten \rangle \right\}
\ee
Where $a$ and $b$ are some suitable constants and $m$ is mass of hyper- 
multiplets. $k$ and $l$ are determined by indices of some elliptic operator. 
$\langle massive \;\;Topological \;\; QCD \rangle $
is a vacuum expectation value 
of an observable with the action of massless Topological QCD.
$\langle Donaldson \rangle$ 
stands for vacuum expectation value 
of an observable with the action of Donaldson-Witten theory \cite{DW},
and is called Donaldson invariants.
$\langle Seiberg-Witten \rangle $ 
stands for vacuum expectation value of an observable with the action of 
Abelian Seiberg-Witten topological field theory \cite{TQCD},
and is called Seiberg-Witten invariants.
The left hand side of the above equation is regular in 
the massless limit, $ m \rightarrow 0$, 
so the Donaldson part and Seiberg-Witten part in the right hand side 
have to cancel each other.
In this theory, mass terms lead a relation between vacuum expectation value 
of Higgs and matter fields.
All computations in this paper are done in a weak coupling limit (or large 
scaling limit).
 If the weak-strong 
duality relation is necessary for understanding the relation
 of Donaldson invariants and Seiberg-Witten invariants, 
 it is natural to think that
massive particles decouple in the weak coupling limit as Witten 
mentioned in 
\cite{W-1} \cite{W-3}
  .
But, in the proof of Hyun-J.Park-J.S.Park mass terms do not decouple
, and  the duality relation is not used.  
Mathematicians did not use the duality relation similarly in their proofs
 \cite{tyurin} \cite{ot}.
This fact implies that mass terms of matter fields do not play 
 essential roles 
in the relation between Seiberg-Witten and Donaldson invariants .
The only important thing is to separate the path-integral into the
Donaldson's irreducible part and Seiberg-Witten's 
reducible part in their theory.

In this paper, we investigate reducibility of the gauge 
connections in massless
Topological QCD. There are two purposes. The first one is that 
we cull the Abelian Seiberg-Witten part from massless Topological
QCD
without 
Higgs mechanism and  weak-strong duality relation.
The Abelian Seiberg-Witten part appear in massless Topological QCD
as reducible connection part.
The second purpose is to obtain the new relations of 
massless Topological QCD and their topological invariants.
Especially, we will get a result that insists some topological invariants 
which contain the Abelian
Seiberg-Witten invariants. 
This topological invariants is provided as reducible connection part 
of this Topological QCD.
As a result of these relations,
we find that the path-integral from non-Abelian extended 
monopoles 
\cite{TQCD}\cite{ot}\cite{ot2}
is separated into Donaldson
parts and non-Abelian Seiberg-Witten parts.
We use
 regularization for zero mode of a scalar field, which do not 
remove zero-modes but shift them to infinitesimal eigenstates without 
BRS-like SUSY breaking with giving perturbation by external fields. 
As a result of this perturbation, 
we obtain new relations between Seiberg-Witten invariants extended
to non-Abelian gauge and Donaldson Seiberg-Witten invariants.
And some identities of U(1) topological invariants from the 
relation are given from the relation of SUSY symmetry
of the Topological QCD and external fields.

This paper is organized as follows.
We set up massless topological QCD whose action
 do not have Higgs potential, in 
 section 2.
Separation of vacuum expectation value of observables into
reducible connection part and 
irreducible connection part is done in section 3. 
We get identities of Abelian Seiberg-Witten invariants and obtain the 
relation of massless Topological QCD and reducible connection part
in section 4.
We will find new formulas in there.
In the last section, we summarize 
 and discuss our conclusions.

 
\section{ Massless Topological QCD}

In this section we set up massless Topological QCD modified slightly 
to separate a correlation function 
 into a reducible connection part and a irreducible connection part 
and study the relations between non-Abelian and
Abelian Seiberg-Witten theory with no Higgs mechanism. Hence Higgs 
potential like $[ \pbar , \p]^2$ do not appear here.
We will find later in this section how the Donaldson invariants 
are embedded in massless 
Topological QCD.

 Topological QCD were already constructed by Hyun-J.Park-J.S.Park
J.M.F.
Labastida and ${\rm Mari\tilde{n}o}$
by twisting N=2 SUSY QCD 
\cite{TQCD}
.
Donaldson theory and Seiberg-Witten theory are
 analyzed as topological field theory in references 
\cite{DW}\cite{TQCD}\cite{P.R.}\cite{HPP}
.
Basically we use the Hyun-J.Park-J.S.Park theory and notation in 
\cite{HPP}
.
In the following, we only consider SU(2) gauge
group and 4-dimensional
compact Rieman manifolds with $b^{+}_{2}>2$ as a back ground manifold.

The action is 

\be
S_{QCD}=-\brs V
\ee
where 
\be 
V=& & \intg [{\chi^{\mu \nu}}_a(H^{a}_{\mu \nu}-i(\mono)) \\
&-& \frac{1}{2} g^{\mu \nu}(\dm \pbar)_a {\lambda}^{a}_{\nu} +(X_{\qtil}^{\al}
{\psi}_{q \al}-{\psi}_{\qtil}^{\al} X_{q \al})] \nonumber
\ee
and under SUSY(BRS like) transformations, Yang-Mills fields 
are transformed as 

\be
\begin{array}{clr}
 \brs A_{\mu} = i{\lambda}_{\mu}, & \brs \chi_{\mu \nu} =H_{\mu \nu}, &
 \brs \pbar =i\eta, \\
 \brs {\lambda}_{\mu}=-\dm \p, & \brs H_{\mu \nu}=i[\p ,\chi_{\mu \nu}], & 
 \brs {\eta}=[\p ,\pbar ], \\
 \brs  \p=0,  & & 
\end{array}
\ee
and the matter fields are transformed as, 

\be 
\brs q^{\ald}&=&- {\psib}^{\ald}_{\qtil}, \; \; \; \; \; \;
\brs {\psib}^{\ald}_{\qtil}\; =\; -i{\p}^a T_a q^{\ald}, \nonumber \\
\brs {\qt}_{\ald}&=& -{\psib}_{q \ald}, \; \; \; \; \; \;
\brs {\psib}_{q \ald}\;=\;i{\qt}_{\ald} {\p}^a T_a  ,   \nonumber \\
\brs {\psi}_{q \al}&=& -i{\s}^{\mu}_{\al \ald} \dm q^{\ald} + X_{q \al} ,
\nonumber \\
\brs X_{q \al} &=& i{\p}^a T_a {\psi}_{q \al} 
-i{\s}^{\mu}_{\al \ald}\dm {\psib}^{\ald}_{\qtil} 
+{\s}^{\mu}_{\al \ald}{\lambda}_{\mu}^a T_a q^{\ald} \nonumber \\
\brs {\psi}^{\al}_{\qtil} &=&
i\dm {\qt}_{\ald} \bar{\s}^{\mu \ald \al}-X^{\al}_{\qtil}, \nonumber \\
\brs X^{\al}_{\qtil}&=& 
i{\psi}^{\al}_{\qtil} {\p}^a T_a -i\dm{\psib}_{q \ald}\bar{\s}^{\mu \ald \al}
 +{\qt}_{\ald}\bar{\s}^{\mu \ald \al}\lambda_{\mu}^{a} T_a.
\ee
These transformation laws are obtained by the usual way of twisting 
N=2 SUSY QCD. 
The matter fields $q$ are sections of 
$W^{+}_{c} \otimes E$  where  $W^{+}_{c} $ is $spin^c$ bundle 
and $E$ is a vector bundle whose fiber is a vector space
of a representation of the gauge group $SU(2)$.
The topological action is given after integrating out the
auxiliary fields $H_{\mu \nu} , X_q$ and $X_{\qtil}$ as
\be
&S_{QCD}&=\intg \left( \right. \frac{1}{4}
{\left| \mono \right|}^2 +\frac{1}{2}{\left| \s^{\mu} \dm q \right|}^2
 -{\frac{1}{2}}\gmn (\dm \pbar)_a(D_{\nu} \p)^a
\nonumber \\
&&- \frac{i}{2} \gmn [{\lambda}_{\mu},\pbar ]_a {{\lambda}^a}_{\nu}
-i{{\chi}^{\mu \nu}}_a [\p ,\chi_{\mu \nu}]^a 
+{\chi}^{\mu \nu}_a {(d_A \lambda)^{+a}}_{\mu \nu}
+\frac{i}{2} \gmn (\dm \eta )_a {{\lambda}^a}_{\nu}
 \nonumber \\
&&
-i{{\chi}^{\mu \nu}}_a {\psib}_q {\bar{\s}}_{\mu \nu} T^a q
+i{{\chi}^{\mu \nu}}_a  \qt {\bar{\s}}_{\mu \nu} T^a {\psib}_{\qtil}
-i\dm {\psib}_{q \ald} {\bar{\s}}^{\mu \ald \al} {\psi}_{q \al}
-i{{\psi}_{\qtil}}^{\al}{{\s}}^{\mu}_{\al \ald }
\dm {\bar{\psi}}_{\qtil}^{\ald} \nonumber \\
&&
+2i{{\psi}_{\qtil}}^{\al} {\p}_a T^a {\psi}_{q \al}
+{\qt}_{\ald} {\lambda}_{\mu a} T^a {\bar{\s}}^{\mu \ald \al} {\psi}_{q \al} 
+{{\psi}_{\qtil}}^{\al}{{\s}}_{\mu \al \ald}{\lambda}_{\mu a}
T^a q^{\ald}\left.\right).
\ee
Where the indices $\al$ and $\ald$ are omitted and we do not change
the position of these indices to keep the sign of each terms in 
the following.
This action is constructed in order to lead the most important 
fixed points,

\be 
\mono =0 , \; \; \; {\s}^{\mu}\dm q=0.
\ee
The Eqs.(7) are monopole equations extended to non-abelian gauge group
and often they are called non-abelian Seiberg-Witten 
monopoles \cite{TQCD}\cite{ot2}. \\

Usually, Seiberg-Witten theory has a Higgs potential $ [\p , \pbar ]^2 $.
Spontaneous symmetry breakdown occur if the vacuum expectation
value of Higgs fields $ \left< \p \right>$ is non-zero. 
Then we get $U(1)$ monopole(Seiberg-Witten) equations.
But one of our purposes is to clarify whether relations
of topological invariants can be understood without 
Higgs mechanism and  weak-strong duality relations. So the 
Higgs potential is not included in our theory. 
(But if we add these potential, 
they do not disturb following discussion.)\\

Later in this section, we find how the Donaldson invariants
are embedded in this theory.
We study three kind of fixed points, which give important contribution
to vacuum expectation value.
The path-integral is expressed as a sum of three parts.
We call a part of the path-integral from fixed points determined by $F^{+}=0$
and $q=0$ as Donaldson part.
We denote a part whose gauge connections $A$ of fixed points
determined by Eqs.(7) are reducible as Abelian Seiberg-Witten part 
and a part which has irreducible connections at the fixed point 
as Non-Abelian Seiberg-Witten part.
The observable is 

\be
\exp \left(  \frac{1}{4{\pi}^2}{\int}_{\gamma} 
Tr \left( i\p F + \frac{1}{2} \lambda \wedge
\lambda \right) - \frac{1}{8{\pi}^2} Tr {\p}^2 \right),
\ee  
where $\gamma$ is in a 2-dimensional homology class i.e. 
$\gamma \in H_{2}(M;{\bf Z})$.
Now let us separates vacuum expectation value with Donaldson invariants
 from non-Abelian Seiberg-Witten invariants.
If fixed points $\langle q \rangle $ 
from Eqs.(7) is zero then the contribution to the expectation 
value of  (8) is from only 
Donaldson theory because now our fixed point equation is 
simply as
\be
\Fp_{\mn}=0.
\ee
We know that we can estimate exactly the vacuum expectation values 
of this observables 
by one-loop approximation around the fixed point determined 
by the Eqs.(7)
\cite{P.R.}\cite{W-2}.
We can decompose the action of Eq.(6) into two parts \cite{HPP},as
\be
S_{QCD}=S_{D}+S_{M},
\ee
where $S_{D}$ is action of Donaldson-Witten theory
\cite{DW},
\be
S_{D}&=&\intg Tr \left[ \right.
\frac{1}{4} \Fp_{\mn} \Fp{}^{\mn} -\frac{1}{2} \gmn \dn{\pbar} \D 
-i\x^{\mn}  \left[ \p , \x_{\mn} \right]
\nonumber \\
& &
+\x^{\mn}\left( d_{A} \la \right)^{+} 
+\frac{i}{2} \gmn \left( \dm \eta \right) \la_{\n} 
-\frac{i}{2} \gmn \left[ \la ,\pbar \right] \la_{\n} \left. \right]
\ee
and $S_{M}$ is matter part,

We find that the quadratic part of the matter part action $S_{M}$ 
 is given by
\be
S_{M}^{(2)}&=&\intg \left[ \right. -3X^{\al}_{\qtil}X_{q \al} +iX^{\al}_{\qtil}
{\s}^{\m}_{\al \ald} \dm q^{\ald}
+i \dm {\qt}_{\ald} {\sbar}^{\m \ald \al} X_{q \al}\nonumber \\
& &
-i {\dm {\psib}_{q \ald}}{\sbar}^{\m \ald \al} \psi_{q \al} 
-i  {\psi_{\qtil}}^{ \al}{\sbar}^{\m}_{\al \ald}{\dm {\psib}_{\qtil}^{ \ald}}
\left. \right] .
\ee
Note that the gauge field $A_{\m}$ used in $\dm$ is an external field.
We expand the gauge field around the solution of Eqs.(7) denoted as $A_c$, 
, as 
\be
A=A_c +A_q.
\ee
,where $A_q$ is a quantum fluctuation around the $A_c$. 
So the covariant derivative in the $S_{M}^{(2)}$
is written by external fields $A_c$ as $ d_A = d + A_c $.
Note that $A_c$ is a irreducible connection because
we set $b^{+}_2 \geq 1$.
After the Gaussian integrals of $X,q,\psi$ and $\psib_q$
, from Eq.(12) we get
\be 
{det\left( -\pi \right) }_{W^{-}} {det\left( -4\pi \right)}_{W^{+}}.
\ee
The indices $ W^{-}$ and $W^{+} $ of determinants (14) show that
the determinants are defined in the subspaces of 
$ W^{-},W^{+} $, and we use the similar notations also
in the following.
Therefore we can understand that the Donaldson invariants appear with 
the determinants (14) in massless Topological QCD and this fact 
is used in section 4.

In the next step, we want to evaluate the other part which 
correspond to the Seiberg-Witten theory and separate the path-integral 
into a reducible connection part and an irreducible part.
But, as we will see it soon, 
it is impossible to separate the Seiberg-Witten part into 
the non-Abelian Seiberg-Witten brunch and the Abelian Seiberg-Witten 
brunch in the same manner as above. 
In this case fixed points value $ \left< q \right>$ is non zero.
We consider the SU(2) gauge group. Our massless theory has 
no spontaneous symmetry break down, so we have to treat separately
reducible gauge connections and irreducible connection  by some way.
Judgments whether the connections are irreducible or not can
be done by examination of the existence of $\dm$ zero-mode.
(See appendix A.)
Strictly speaking, the following two propositions are same.
\begin{itemize}
\item{$d_{A}$ : $Ad\; \xi \bigotimes {\bigwedge}_{0} \rightarrow 
 Ad \;\xi \bigotimes {\bigwedge}_{1} $ is injection.}
\item{$A$ is a irreducible connection.}
\end{itemize}
Where we represent a vector bundle with a structure group SU(2) as $\xi$.
So we use this condition to divide the contribution to 
vacuum expectations into one from Abelian Seiberg-Witten theory
and another from non-abelian Seiberg-Witten thery.
At first, we pay attention to
$ \pbar $ equation,
\be 
-\frac{1}{2} {D^{\mu}} \D - \frac{i}{2} \left[ {\la}_{\m},{\la}^{\m} \right]=0.
\ee
 $\la$ in this equation is determined by $\eta$ and $\chi$ equations
like appendix B or references 
\cite{P.R.}
.
When there is no zero mode solution of $\la $ then this equation is a 
distinction formula
of reducibility i.e. $-\frac{1}{2} {D^{\mu}} \D =0$.
 So we can conclude that the connection is reducible 
and the contribution to vacuum expectation value is from the U(1) abelian
Seiberg-Witten theory when $ \left< \p \right> \neq 0 $.
But unfortunately we can not distinguish connection by this method 
when there are $\la $ zero mode solutions.
In the next section we find new approach to separate the contribution of
abelian Seiberg-Witten branch
from non-Abelian Seiberg-Witten part.



\section{Separation of reducible connection part}
In this section we construct a new theory in which 
vacuum expectation value are separated into three parts i.e. Donaldson 
, Abelian Seiberg-Witten and non-Abelian Seiberg-Witten part.
We use the determinant obtained by integration of $\p$ and $\pbar$ 
to find whether connections are reducible or not.
 Therefore we consider the case
when the determinant vanishes. Usually we avoid this case and remove
zero eigenvalue states with various ways.
For example in \cite{B.B}, zero-modes give a symmetry to whole action 
and they are removed by BRS method. 
In our case this method can not be used because zero-mode dose not give 
the existence of a local symmetry.
But in this section we take zero-modes into account 
, and we find that this zero-modes play a essential role
to distinguish between reducible connections and irreducible connections.

  In the same way as section 2, we pay attention to Eq.(15).
From Eq.(15), we get the vacuum expectation value of 
the scalar fields $ \p$ in 1-loop order as
\be
\left< \p \right> = -\frac{i}{D^{\m} \dm} \left[ \la_{\n} , \la^{\n} \right]
\ee

Now we have to recall that if ${b}^{+}_{2} >0$, anti-selfdual
 connections which satisfy 
the equation (9) do not contain the reducible connections with U(1)
isotoropy group
\cite{D-1}\cite{Uh}
, but the connections which satisfy the 
monopole equations (7) contain such reducible connections in general.
When the connection is reducible then $\dm$ has zero-mode on 0-form,
(see the appendix A).
Then the Green function $\frac{1}{D^{\m} \dm}$ has singularities.
Normally we avoid this kind of singularity to define a meaningful theory.
However, in the present case, this singularity pays the important role.
Reducibility of the gauge connection is judged by the zero-mode.
The theory should keep holding characteristic properties of zero-modes 
while it is regularized.
Such problem doesn't exist in the Donaldson theory.
Or we can say that we set the condition ${b}^{+}_{2} >2$ to avoid 
the complication of reducible connections in Donaldson theory.
But in the Seiberg-Witten theory it is the most important merit 
that the gauge group is U(1) Abelian group. So, if we take away 
such reducible connections, then this Topological QCD has almost 
no value.
Therefore we have to manage the singularities in the Green function.
Usually we dispose of singularities of this kind
by removing zero-modes, inducing mass terms and so on.
The following way makes it possible.

Before considering regularization we ascertain that 
these singularities make topological symmetry break.
To see it concretely, we introduce a BRS exact observable, 
$\brs \left( \pbar \la_{\m} \right)$.
If there are no singularities, topological symmetry is 
not broken and vacuum expectation value of this observable
vanishes.
But as we saw before, the propagator $ \left< \pbar \p \right> \sim
\frac{1}{D^{\m} \dm}$ is singular at least in tree level. 
To avoid these singularities we 
add regularization term $i\ep \pbar \p $ to our Lagrangian and 
the propagator is changed to $\frac{1}{D^{\m} \dm -i\ep} $, naively.
Because the $ \pbar \p $   is not 
invariant under the BRS-like SUSY transformation (4)(5),
 the vacuum expectation 
value is 
\be
\left< \brs \left( \pbar \la_{\m} \right) \right> &=&
\intD \brs \left( \pbar \la_{\m} \right) e^{-S} \nonumber \\
&=& \ep \intD \left( \pbar \la_{\m} \right) 
\brs \left( \pbar \la_{\m} \right) e^{-S} \nonumber \\
&=& \ep \left< \pbar \la_{\m} \eta \p \right> \nonumber \\
&=& \ep \left( \left< \pbar \p \right> \left< \la_{\m} \eta \right> 
+ \cdots \; \; \; \right).
\ee
(See references 
\cite{sako}.)
The last equality is from the Wick's theorem.
It make clear that the singularities of the propagator 
$\left< \pbar \p \right> \simeq \frac{1}{\ep} $
 cause the right hand side of Eq.(7) 
non-zero in a limit as $\ep \rightarrow 0$.
This fact means the vacuum expectation value
of the BRS exact observable is non-zero.
This is topological symmetry breaking.
It is equivalent to a phenomenon observed in
\cite{sako}\cite{Z-L}\cite{Fu}.
This is seen after integration by $ \pbar$. We get a delta function, 
\be
\delta \left( -\frac{1}{2} {D^{\mu}} \D - \frac{i}{2}
 \left[ {\la}_{\m},{\la}^{\m} \right] +i\ep \p \right)
= \sum_{{\rm all}\;{\hat{\p}}_{\ep} }
\frac{ \delta \left( \p -{\hat{\p}}_{\ep} \right)}{\left| 
det\left( D^{\m} \dm -i\ep \right) \right| },
\ee
where ${\hat{\p}}_{\ep}$ is the solutions of  
$-\frac{1}{2} {D^{\mu}} \D - \frac{i}{2}
 \left[ {\la}_{\m},{\la}^{\m} \right] +i\ep \p =0$ and $\la$
is a zero-mode which is determined by $\eta $ and $\x$ equations. 
(see the Appendix B).
We know it from seeing the delta function of (18) that the determinant 
${\left| det\left( D^{\m} \dm -i\ep \right) \right| }$
is zero in a limit as $\ep$ approaches zero when the connection is reducible, 
and these singularities break the topological symmetry.
From this consideration, it seems that the regularization
term is not suitable. For our purpose we do not hope topological symmetry
breaking, so we have to choose the 
regularization term to be invariant under BRS-like
SUSY transformations. \\

Here we define the determinant 
by adding infinitesimal sift terms to our Lagrangian.
Our problem is that we could not separate 
vacuum expectation value into a reducible 
connection part and a irreducible connection part
with the way in the previous section.
In this section, the method used in 
\cite{sako}
adapt to the separation.
We pay attention to the determinant
$det\left( D^{\m} \dm \right)$ which is obtained by integration 
of $\p$ and $ \pbar $.
We saw above that vanishing of this determinant cause topological
symmetry breaking or at least cause some singularity.
Usually we used to modify determinants  
by removing zero eigenvalues.
But these zero-modes are necessary to judge the reducibility of the gauge connection.
So we do not remove but shift them by infinitesimal perturbation.
This infinitesimal perturbation is given by adding some shift terms to our Lagrangian. 
We denote this shift term as $\pbar f_{\ep}$ where $f_{\ep}$ 
is a some functional and $f_{\ep}=0$ as the limit of ${\ep \rightarrow 0}$.
Similarly we also shift the determinant 
which is obtained by $\eta$,$\chi$ and $\la$ integrations.(See the appendix
B.)
This shift term is represented to $\eta \zeta_{\ep}$ in the following where 
$\zeta_{\ep}$ is a some fermionic functional and 
$\zeta_{\ep}$ vanishes in the limit, ${\ep \rightarrow 0}$.
We mention a little more about this shift.
The 
$\eta$  integral
of  $\gmn \left( \dm \eta \right) \la_{\n}$ in action (6) vanishes,
if the covariant derivative $\dm$ acting to the $\eta$ has 
zero-modes.(Note that $\la_{\mu}$ zero-modes define the dimension
of the moduli space, but $\eta$ zero-modes do not have such a roles in 
usual case.)
Usually we take away this zero-modes, but now instead of doing so 
we add infinitesimal shift terms
$\zeta_{\ep}$ to the Lagrangian, in order 
to shift the fermionic zero-eigenstates into infinitesimal eigenvalue states.
Total Lagrangian is now written as
\be
S=S_{QCD} + \intg (\pbar f_{\ep} +\eta \zeta_{\ep})
\ee

Here let us enumerate conditions for the infinitesimal terms.
\begin{quote}
(a) Shift terms are invariant under SUSY transformations (4) and (5).
\end{quote}
This condition is necessary to avoid topological symmetry breaking
as we saw it above.
\begin{quote}
(b) It is desirable that non-zero solutions,
$\p$ and $\la$, of the following $\pbar$ and $\eta$ equations 
do not vanish by
adding shift terms:
\end{quote}
\be
-\frac{1}{2} {D^{\mu}} \D - 
\frac{i}{2} \left[ {\la}_{\m},{\la}^{\m} \right]
- f_{\ep}=0 \\
\frac{i}{2} \dm {\la}^{\m} -\zeta_{\ep}=0.
\ee
From this condition, observable contain non-trivial one.
The $\la$ in Eq.(20) is determined by Eq.(21) and fermionic
field equations, 
(refer \cite{P.R.} and see appendix B).
When $f_{\ep}$ contains the field $\p$ like
\be
f_{\ep}= \ep g\p +\ep h
\ee
where $g$ and $h$ are some functionals which do not contain 
scalar field $\p$,
$\left< \p \right>$ is represented as 
\be
\left< \p \right> = 
-\frac{1}{D^{\m} \dm +\ep g}\left( i \left[ \la_{\n} , \la^{\n} \right]
+ 2 \ep h \right).
\ee 
Hence $\left< \p \right>$ becomes 
order $\epsilon$ i.e.$\left< \p \right> = 
-\frac{1}{D^{\m} \dm +\ep g}\left(
+ 2 \ep h \right) =O\left( \ep \right)$
if there is only $\la=0$ solution of (21) and $D_{\m}$ has no zero-mode.
Then the vacuum expectation of observable (8) is 
mere $1 +O\left( \ep \right)$ and this is not suitable.
Note that it is not desirable that $f$ is independent from $\p$.
This reason is made clear by next condition (c).
So we can not put $g$ zero in Eq.(22).
The righthand side of Eq.(22) has at most linear in the
field $\p$, however  $f_{\ep}$ may contain higher power terms on $\p$.
For simplicity, we treat only case Eq.(22).
The 3rd condition for infinitesimal shift terms is
\begin{quote}
(c) following two operators have no zero-modes regardless of whether 
gauge connections are reducible or not.
\end{quote}
\be
\frac{1}{2} D^{\m} \dm +
\ep \frac{\delta f}{\delta \p} \equiv \frac{1}{2} \dse .\\
\frac{i}{2} \dm  
-\epd \frac{\delta \zeta}{\delta \la^{\m}} \equiv \frac{i}{2} \de.
\ee
Where the operators (24) and (25) operate on $Ad(E)$ valued 
0-form. 
This condition is just to shift determinants from zero
to infinitesimal finiteness when gauge connections are 
reducible and $\dm$ has zero-modes.
From these conditions, we can take count of $\eta$ and $\p$ zero-modes 
when calculate $det \dse $ obtained by
$\p $ and $\pbar$ integrations and $det T_{\ep}$
(see appendix.B) whose $T_{\ep}$ has $\de $ in first row and is
obtained by fermionic fields integrations.

Example of the additional terms $\pbar f_{\ep}$ and $\eta_a \zeta_{\ep}^a$
which satisfy the conditions (a)(b)(c) are 
\be
\pbar f_{\ep}+\eta_a \zeta_{\ep}^a &=&
\brs \left( \epd \pbar_a {{\la}^a}_{\m} m^{\m} + \ep n(\qt \pbar \psib_{\qtil}-
\psib_{q} \pbar q) \right)
 \nonumber \\
&=& \epd \pbar_{a} \left( \dm \p \right)^a m^{\m} + \ep
\pbar_{a} \left( n \qt (\p  T^a +T^a \p  ) q -2in \psib_{q} T^a \psib_{\qtil} 
\right) \nonumber \\
& &+\epd \eta_{a} \left( {{\la}^a}_{\m} m^{\m} \right)- \ep \eta_{a} \left(
n \psib_{q} T^a q + n \qt T^a 
\psib_{\qtil} \right),
\ee
where $m_{\m}$ and $n$ are some back ground fields which are chosen to satisfy 
the conditions (b) and (c) and are gauge singlet. These external 
fields do not break SU(2) gauge symmetry and topological symmetry 
because they are gauge singlet fields.
Infinitesimal constant number $\ep$ and $\epd$ are
independent each other.
 We can add shift terms (26) to our action since
(26) are BRS-like SUSY invariant and have 0 U-number.
Note that sometime the determinant $ det \dm D^{\m} $ obtained 
by $\p$ and $\pbar$ integrations is ignored in other papers 
because this determinant can be countervailed by a Fadeev-Popov
determinant of the SU(2) gauge fixing. But in our theory, 
this determinant plays a essential role for separating reducible connections.

Next we consider the zero limit of $\ep$ and $\epd$.
We take this limit after functional integrals.
When we calculate around fixed points of 
irreducible connections which have no 
zero modes of $\dm $, additional terms like (26) play no role
and there is no modification in Donaldson and non-Abelian Seiberg-
Witten theory in the limit as $\ep$ approaches zero.
But if a connection of a fixed point is reducible some different
points appear.
At first, we can estimate the order of the delta function obtained
by $\pbar$ integrations as
\be
{\delta} \left( -\frac{1}{2} {D^{\mu}} \D - \frac{i}{2}
 \left[ {\la}_{\m},{\la}^{\m} \right] +  f_{\ep} \right)
= \sum_{{\hat{\p}}_{\ep} }
\frac{ {\large \delta} \left( \p -{\hat{\p}}_{\ep} \right)}{\left| 
det\left( D^{\m} \dm -\frac{\delta f_{\ep}}{\delta \p}
 \right) \right| } \nonumber \\
\frac{ 1}{\left| 
det\left( D^{\m} \dm -\frac{\delta f_{\ep}}{\delta \p}
 \right) \right| } \sim \left\{ \begin{array}{c}
1 \ \ ({\rm for \;irreducible\; connections\;} A_{\m}) \\
\ep^{-l} \ \ ({\rm for\; reducible\; connections\;} A_{\m})
\end{array}\right.
\ee
where $l$ is a dimension of 0-cohomology $H^{0}_{A}$ i.e. $l=\dim \;\ker(\dm)$.
We denoted $\p$ that do not make delta function vanish as ${\hat{\p}}_{\ep}$. 
Especially it is important to notice that this determinant can be chosen
to depend on not $\epd$ but $\ep$ like the case (26). Since the
$\epd \pbar_{a} \left( \dm \p \right)^{a} m^{\m} $
term in shift terms (26) vanishes under $\dm$ zero-modes integral.
For infinitesimal $\ep$ and $\epd$, $\epd$ terms in the determinant in Eqs.(27)
are negligible. Note that the shift terms which break the balance of 
bosonic and fermionic shift terms in zero-mode integral
like Eq.(26) have less variations.
In general, $\ep$ or $\epd$ appear in the both coefficients of $\pbar$ term and $\eta$ term
since $\pbar$ is a super partner of $\eta$.

The equations to obtain vacuum expectation value of $\p$ change also as
\be
\lim_{\ep \rightarrow 0}\left< \p \right> 
&=& -\lim_{\ep \rightarrow 0}\frac{i}{D^{\m} \dm -\ep g}
 \left( \left[ \la_{\n} , \la^{\n} \right] + \ep h \right) \nonumber \\
&=& \left\{ \begin{array}{c}
\p_{c} \ \ ({\rm for \;irreducible\; connections\;} A_{\m}) \\
\p_{c} +\ep {\p}' \ \ ({\rm for\; reducible\; connections\;} A_{\m})
\end{array}\right.
\nonumber \\
\ep {\p}'&\equiv&\lim_{\ep \rightarrow 0}\frac{i\ep}{D^{\m} \dm -\ep g}
( g\p_{c} + h).
\ee
Where $\p_{c}$ is a solution of the equation (15) 
of each connections $A_{\m}$.
Note that $\ep {\p}'$ may be finite in the 0-limit of $\ep$ 
since $D_{\m}$ has zero-modes. 

In a similar manner, we estimate a delta function which obtained by
fermionic field $\eta$ integration as 
\be
\prod_{\eta_{0}}\zeta_{\ep} \prod_{\eta'}{\large \delta} 
(\frac{i}{2} \dm {\la}^{\m} -\zeta_{\ep})
 \sim \left\{ \begin{array}{c}
\prod_{\eta}{\large \delta} (\frac{i}{2} \dm {\la}^{\m} -\zeta_{\ep})
 \\
 ({\rm for \;irreducible\; connections\;} A_{\m}) \\
O(\ep,\epd)^{l} \prod_{\eta'}{\large \delta} (\frac{i}{2} \dm {\la}^{\m} -\zeta_{\ep})
\\ 
({\rm for\; reducible\; connections\;} A_{\m}).
\end{array}
\right.
\ee
Where $\eta_{0}$ are  $\dm$ zero-modes and $\eta'$ are other
non-zero modes and $O(\ep,\epd)^l$ is order $\ep^n \epd^{m}$ where 
$n+m=l,(n,m=0,1,2,..,l)$.
Or we can say it as follows. The first row of $T_\ep$ that is $\de$
goes back to $\dm$ in the zero limit of $\ep$ and $\epd$
when connections of fixed points are irreducible.
On the other hand when connections of fixed points are reducible,
$\eta$ and $\la$ integral can be written by separation $\dm$ zero-modes
$\eta_{0}$ 
\be
\int {\cal D} \eta' {\cal D}\la \;\; \exp\left( 
-\intg ((\dm \eta'){\la}^{\m} -\zeta_{\ep} \eta') \right)
\int {\cal D} \eta_{0} \;\; \exp\left(
-\intg \eta_{0}\zeta_{\ep}  \right).
\ee
So we can estimate $det \left| T \right|$ as 
\be
\lim_{\ep ,\epd \rightarrow 0} det \left| T_{\ep} \right|=
\left\{ \begin{array}{c}
det \left| T \right| \ \ ({\rm for \;irreducible\; connections\;} A_{\m}) \\
\lim_{\ep ,\epd \rightarrow 0}
(O(\ep,\epd))^l \ \ ({\rm for\; reducible\; connections\;} A_{\m})
\end{array}
\right.
\ee
Where  
$O(\ep,\epd)^l$ is a determinant of a matrix in which 
the $\eta_{0}$ row $\la_{\m}$ column of $T$ is replaced  with 
$\frac{{\delta} \zeta_{\ep}}{\delta \la}$.
In the example (26), 
the $\eta_{0}$ row $\la_{\m}$ column element in $T_\ep$ is
\be
\frac{{\delta} \zeta^a_{\ep}}{\delta \la_{\m b}}
=\epd \delta^{ab} m^{\m} .
\ee

From Eqs.(27) and (29), we conclude an order of 
 vacuum expectation values is given by,
\be
\left\{ \begin{array}{c}
1 \ \ ({\rm for \;irreducible\; connections\;} A_{\m}) \\
\left( \frac{O(\ep,\epd)}{\ep} \right)^{l} \ \ 
({\rm for\; reducible\; connections\;} A_{\m}).
\end{array}
\right.
\ee
We can conclude from (33) that if we set $\ep$ and $\epd$ as same order,
then our path-integral are sums over reducible and irreducible
connection parts with an equal weight.

Let us consider changing the ratio $\epd / \ep$ and changing the contributions
from reducible connection part in our path-integral.
The ratio of $\ep$ and $\epd$ can be changed without
changing of vacuum expectation value. This fact is seen as follows.
We put $\epd$ as $\epd = k\ep$ and $k$ is some positive real number.
When an action is given as
\be
S=\intg \brs V + \ep \brs F + \epd \brs G  
\ee
where $V,F$ and G are any 
functionals, then the change in a
vacuum expectation value of any observable $O$
which satisfy $\brs O =0$ under an infinitesimal deformation of $k$
is 
\be
\frac{\delta}{\delta k} \left< O \right> &= &
\intD \; O \;\frac{\delta }{\delta k}
\left( \epd \brs G \right) e^{-S} \nonumber \\
&=& \intD \;\brs \left( O (\ep G) \right) e^{-S} =0.
\ee
So we can change $k$ without changing vacuum expectation value.
In our case, only reducible connection part depend on $k$.
We denote $\left< O \right>_{IR}$ as a irreducible connection (non-abelian connection)
part of $ \left< O \right>$ and denote $\left< O \right>_{R}$ as a reducible 
connection (Abelian connection) part, then the only $\left< O \right>_{R}$ depend 
on the ratio $k$. When the power expansion of $k$ of $\left< O \right>_{R}$
is written 
as $\left< O \right>_{R}=\sum_{n=0}^{l} \left< O \right>_{R,n} \; k^n$, 
vacuum expectation value $\left< O \right>$ is expressed as
\be
\left< O \right> &=& \left< O \right>_{IR}+ \left< O \right>_{R}
\nonumber \\
&=& \left< O \right>_{IR}+ \sum_{n=0}^{l} \left< O \right>_{R,n}  \; k^n
\ee
Since $\left< O \right>$ is $k$ independent, 
we obtain 
\be
\left< O \right>_{R,n}  = 
\left( \frac{\delta}{\delta k} \right)^n \left< O \right>_{R}\left. \right|_{k=0}
=0,\;\;(n=1,...,l).
\ee
This fact means that
we can remove the contribution to vacuum expectation value from
reducible connection without a $k^0$ proportional term $\left< O \right>_{R,0} $.
Note that $\left< O \right>_{R,0}= \left< O \right>_{R} \left. \right|_{\epd=0}$.
This is the most important
fact to derive the relation of Abelian Seiberg-Witten invariants
in the next section.
Note that the vacuum expectation values $\left< O \right>$
is invariant under changing of $\ep$. 
This is easily ascertained by a similar way of (35).

In the next section, we explicitely investigate a relation of the
non-Abelian Seiberg-Witten, Donaldson and the Abelian Seiberg-Witten
invariants with the information obtained in this section.


\section{The relation of Topological Invariants}

In the previous sections, we prepared the tools for investigation of relations
of the topological invariants i.e. Donaldson, Abelian Seiberg-Witten 
and non-Abelian Seiberg-Witten invariants.
Now we actually construct the formulas of these invariants.
We treat three parts of vacuum expectation value of observable (8) separately.
The first part is Donaldson part which is defined as fixed point $ q=0 $
and gauge connections of the fixed points are irreducible connections.
There is no solution of the instanton equation when 
$b^+_2 \geq 1$ and connections are reducible.
The second part is Abelian part whose fixed points $q \neq 0$ and 
connections are reducible.
The third part is non-Abelian part 
that has irreducible connections and $q \neq 0$.
In our theory, vacuum expectation value of $O$ is defined by
\be
\langle O \rangle &=& \lim_{\ep , \epd \rightarrow 0} 
\intD \;\; O\; \exp(-S) \\
&=& \left\{ 
\begin{array}{c}
\intD \;\; O \;\exp(-S_{QCD}) \nonumber \\
({\rm for\;Donaldson\;and\;non-Abelian\;Seiberg-Witten\;part}) \nonumber \\
 \nonumber \\
\lim_{\ep , \epd \rightarrow 0} \intD \;\; O\; 
\exp({-S_{QCD} - \intg (\pbar f_{\ep} +\eta \zeta_{\ep})})
\nonumber \\
({\rm for\;Abelian\;Seiberg-Witten\;part}).
\end{array}
\right.
\ee
The fact that the $\ep$ terms do not influence irreducible connection part
was seen in previous section.
In noting this point we advance analysis in the following.

{ \bf Donaldson part} 

We already saw the transvers path-integral of this part in section2.
We represent common factors from $H_{\m \n}$ and $X$ integrations
as ${\cal N}$.
Then we can write Donaldson part as,
\be 
{\cal N}det(-4 \pi ) \langle \exp (\tilde{v} + \tau \tilde{u}) {\rangle}_{D}
\ee
where
\be
\tilde{v} &=& \frac{1}{4{\pi}^2}{\int}_{\gamma} 
Tr \left( i\p F + \frac{1}{2} \lambda \wedge
\lambda \right) \nonumber \\
\tilde{u} &=& - \frac{1}{8{\pi}^2} Tr \left({\p}^2 \right)
\ee  
and $\langle O {\rangle}_{D}$ means vacuum expectation value of $O$
 with the 
action (11) of Donaldson-Witten theory.
$\tau$ is a parameter.

{ \bf Abelian Seiberg-Witten part}

In this part, $A_{\m}$ on fixed points are reducible connections.
When we calculate in the large scaling limit it is possible that 
we choose back ground fields $A_c$ in (13) as U(1) connections.
The vector bundle $E$ reduces to the sum of line bundle 
i.e. $E= \zeta \oplus {\zeta}^{-1}$.(See the Appendix A.)
We set the direction of back ground gauge fields to $T_{3}$
i.e. $A_{c \m}=A_{3 \m}T_{3}$.  
Since we take $q$ as fundamental representation of SU(2), we can write 
$q$ as 
\be
q^{\ald}= \left(
{\begin{array}{c}
q^{\ald}_{1} \\
q^{\ald}_{2}
\end{array}} \right).
\ee
We can reduce line bundle $\pmu \zeta$ to $+\zeta$ and $q^{\ald}_{2}=0$
because there is a symmetry defined by $A_{3} \rightarrow -A_{3}$
and $q_{1} \rightarrow q_{2}$ in this part.
So we can estimate the vacuum expectation value of (38)
around fixed points $A_{c}$ and $q_c$ by considering quadratic action
of quantum fields $A$,$q$ and others. 
We expand $A$ and $q$ as
\be
A=A_c +A, \;\;\; \; q=q_c+q
\ee
where the each second term of right hand side is a quantum field
and $A_c$and $q_c$ is chosen as
\be
A_{\m c} =A_{\m 3}T_3 , \; \;\; 
q^{\ald}_c= \left(
{\begin{array}{c}
q^{\ald}_{1} \\
0
\end{array}} \right).
\ee
Next we decompose the action $S_{QCD}$ into two parts as we did in section 2
as 
\be
S_{QCD}\approx S_c +S_t,
\ee
where $S_c$ is the action which consists of the Caltan part of adjoint fields 
and the first component of fundamental representation fields. 
The $S_c$ is written as
\be
S_c &=& \intg \left\{ \right.
\frac{1}{4} \left| F^{+}_{\m \n 3}+\frac{1}{2}\qt_{1} \sbar_{\mn} q_{1} \right|^{2}
+\frac{1}{2} \left| \Ds q_{1} \right|^{2}
-\frac{1}{2} \pmuu \pbar_{3} \pmu \p_{3}
+\x^{\mn}_{3} (\pmuu \la_{\n 3})^{+} \nonumber \\
& &+\frac{i}{2}(\pmuu \eta_{3}) \la^{\m}_{3}
-\frac{i}{2}\x^{\mn}_{3} \psib_{q 1} \sbar_{\mn} q_{1}
+\frac{i}{2}\x^{\mn}_{3} \qt_{1} \sbar_{\mn} \psib_{\qtil 1}
-i( \Ds \psib_{q 1})\psi_{q 1} \nonumber \\
& &-i\psi_{\qtil 1}\Ds \psib_{\qtil 1}
+\frac{1}{2} \qt_{1} \la_{\m 3} \sbar^{\m} \psi_{q 1}
+\frac{1}{2} \psi_{\qtil 1} \s_{\m} \la^{\m}_3 q_{1} \left.\right\},
\ee
where $\Dm = \pmuu - i\frac{1}{2}A_{\m 3}$ and spinor indices $\al$ and $\ald$ 
are omitted.
This action $S_c$ is surely the action of topological Abelian Seiberg-Witten
theory \cite{HPP}\cite{TQCD}.
This Caltan part action is Witten type topological action as follows
\be 
S_c = &\brs_c & [ \frac{1}{2} \x^{\mn}_{3} 
((F^{+}_{3 \mn}+ \frac{1}{2} \qt_1 \sbar q_1)-H_{3\mn}) 
\nonumber \\
& &-\frac{1}{2}(\pmu \pbar_3) \la_{3}^{\mu}  
+(X_{\qtil 1}\psi_{q1} - \psi_{\qtil 1}X_{q1})],
\ee
where we define the BRS-like SUSY as 
\be
\begin{array}{clr}
 \brs_c A_{3\mu} = i{\lambda}_{3\mu}, & \brs_c \chi_{3\mu \nu} =H_{3\mu \nu}, &
 \brs_c \pbar_3 =i\eta_3, \nonumber \\
 \brs_c {\lambda}_{3\mu}=-\pmu \p_3, & \brs_c H_{3\mu \nu}=0, & 
 \brs_c {\eta}=0, \\
 \brs_c  \p=0.  & & 
\end{array}
\ee
\be
\brs_c q^{\ald}_1&=&- {\psib}^{\ald}_{\qtil 1}, \; \; \; \; \; \;
\brs_c {\psib}^{\ald}_{\qtil 1}\; =\; -\frac{i}{2}{\p}^a_3 q^{\ald}_1, \nonumber \\
\brs_c {\qt}_{\ald 1}&=& -{\psib}_{q 1 \ald}, \; \; \; \; \; \;
\brs_c {\psib}_{q 1\ald}\;=\;\frac{i}{2}{\qt}_{1\ald } {\p}_3  ,   \nonumber \\
\brs_c {\psi}_{q1 \al}&=& -{\s}^{\mu}_{\al \ald} \Dm q^{\ald}_1 + X_{q 1\al} ,
\nonumber \\
\brs_c X_{q 1\al} &=& \frac{i}{2}{\p_3} {\psi}_{q 1\al} 
-i{\s}^{\mu}_{\al \ald}\Dm {\psib}^{\ald}_{\qtil 1} 
+\frac{1}{2}{\s}^{\mu}_{\al \ald}{\lambda}_{3\mu} q^{\ald}_1 \nonumber \\
\brs_c {\psi}^{\al}_{\qtil 1} &=&
i\Dm {\qt}_{\ald} \bar{\s}^{\mu \ald \al}-X^{\al}_{\qtil 1}, \nonumber \\
\brs_c X^{\al}_{\qtil 1}&=& 
\frac{i}{2} {\psi}^{\al}_{\qtil 1} {\p}_3 -i\Dm{\psib}_{q 1\ald}\bar{\s}^{\mu \ald \al}
 +\frac{1}{2}{\qt}_{1\ald}\bar{\s}^{\mu \ald \al}\lambda_{3 \mu}.
\ee
Another terms in $S_{QCD}$ have to be expanded around fixed points
$A_{3c}$ and $q_{1 c}$.
The quadratic action of $S_t$ is obtained as  
\be
S_{t}^{(2)}&=&
\intg [ \frac{1}{4}(\DM A_{\n +}-\DN A_{\m +})^{+}
(\DMS A_{\n -}-\DNS A_{\m -})^{+}
\nonumber \\
&+&\frac{1}{8}(\DM A_{\n +}-\DN A_{\m +})^{+}\qt_{2} \sbar^{\mn} q_{1c}
+\frac{1}{8}(\DMS A_{\n -}-\DNS A_{\m -})^{+}\qt_{1c} \sbar^{\mn} q_{2}
\nonumber \\
&+&\frac{1}{4}\left|\qt_{1c} \sbar_{\mn} q_{2} \right|^2
+\frac{1}{2}\left|\Ds^{\ast}q_2 \right|^{2}
-\frac{1}{4}(\Ds^{\ast}q_{2})^{\dagger} \s^{\m} A_{\m +}q_{1c}
-\frac{1}{4}(\s^{\m} A_{\m +}q_{1c})^{\dagger}(\Ds^{\ast}q_{2})
\nonumber \\
&+&\frac{1}{8}\left|\s^{\m} A_{\m +}q_{1c}\right|^2
-\frac{1}{2}(\DM \pbar_{+})(\cD^{\m \ast} \p_{-})
-\frac{1}{2}(\cD_{\m}^{\ast} \pbar_{-})(\cD^{\m} \p_{+})
+\x^{\mn}_{+}(\cD_{\m}^{\ast} \la^{-}_{\n})^{+}
\nonumber \\
&+&\x^{\mn}_{-}(\cD_{\m} \la^{+}_{\n})^{+}
+\frac{i}{2} (\DM \eta_{+})\la^{\m}_{-}
+\frac{i}{2} (\DM^{\ast} \eta_{-})\la^{\m}_{+}
-\frac{i}{2} \x^{\mn}_{-} \psib_{q 2} \sbar_{\mn} q_{1c}
\nonumber \\
&+&\frac{i}{2} \x^{\mn}_{+} \qt_{1c} \sbar_{\mn} \psib_{\qtil 2}
-i(\Ds^{\ast} \psib_{q2})\psi_{q2}
-i\psi_{\qtil 2}(\Ds^{\ast} \psib_{\qtil 2})
\nonumber \\
&+&\frac{1}{2} \qt_{1c} \la_{\m +}\sbar^{\m} \psi_{q2}
+ \frac{1}{2} \psi_{\qtil 2}\sbar^{\m} \la_{\m +} q_{1c} ].
\ee
Where $ \DM =\pmu -iA_{3 \m}$ and we put $T^{\pm}$ as 
$T^{\pm} =\frac{1}{2}(T_{1} \pm i T_{2})$.
$S_{t}^{(2)}$ is transformed with matrices in the same way 
of the appendix B as
\be
S_{t}^{(2)} &=& (A_{\rho -}\; \qt_2) 
\left(
\begin{array}{cc}
M_{A,A} & M_{A,q} \\
M_{q,A} & M_{q,q}
\end{array}
\right)
\left( \begin{array}{c} A_{\tau +} \nonumber \\
q_2 \end{array} \right) \nonumber \\
& & + \pbar_{+}(\frac{1}{2} \DMS \cD^{\m \ast})\p_{-}
+ \pbar_{-}(\frac{1}{2} \DM \cD^{\m})\p_{+}
\nonumber \\
& & +(\eta_{-}\; \eta_{+}\; \x_{0i -}\; 
\x_{0i +}\; \psi_{\qtil 2}\; \psi_{q 2})
\left( {\large T_{t}} \right)
\left( \begin{array}{c} \la_{0 +} \\
\la_{j +} \\
\la_{0 -} \\
\la_{j -} \\
\psib_{\qtil 2} \\
\psib_{q 2} \end{array} \right).
\ee
Where we chose space elements of self-dual field $\x$, i.e. $\x_{0i}$,
as substantial elements.
Matrix $M$ elements are given as follows.
\be
& &( A_{\rho -} \ row \ A_{\tau +} \ column ) \nonumber \\
M_{A,A}&=& -\frac{1}{4}(\DM \delta_{\n \rho} -\DN \delta_{\m \rho})^{+}
(\DM \delta_{\n \tau} -\DN \delta_{\m \tau})
+\frac{1}{8} \qt_{1c} \sbar^{\m} \delta_{\m \rho} \s_{\n \tau}q_{1c}
\nonumber \\
& &( A_{\rho -} \ row \ q_2 \ column ) \nonumber \\
M_{A,q} &=&-\frac{1}{8}(\DM \delta_{\n \rho} -\DN \delta_{\m \rho})^{+}
\qt_{1c} \sbar_{\mn}
-\frac{1}{4}(\qt_{1c} \sbar_{\m}\delta_{\n \rho})\Ds^{\ast} \nonumber \\
& &( \qt_2 \ row \ A_{\tau +} \ column ) \nonumber \\
M_{q,A}&=&\frac{1}{8}\sbar_{\mn}q_{1c}
(\DM \delta_{\n \tau} -\DN \delta_{\m \tau})^{+}
+\frac{1}{4}(\Ds)\s^{\m} \delta_{\m \tau}q_{1c}
\nonumber \\
& &(\qt_2 \ row \ q_2 \ column ) \nonumber \\
M_{q,q} &=&-\frac{1}{2} \Ds \Ds^{\ast} 
+\frac{1}{4} \sbar_{\mn} q_{1c} \qt_{1c} \s^{\mn}.
\ee
Where the index $+$ means anti-selfdual about indices $\m$ and $\n$.
The elements of the matrix $T_{t}$ is obtained 
in the same process in appendix B.
We write it as
\be
\left(
\begin{array}{cccccc}
-\frac{i}{2} \cD_{0} & 0 & -\frac{i}{2} \cD_{j} & 0 & 0 & 0 \\
0 & -\frac{i}{2} \cD^{\ast}_{0} & 0 & -\frac{i}{2} \cD^{\ast}_{j} & 0 & 0 \\
-\frac{1}{2} \cD_{i} & 0 & \frac{1}{2}({\cal PD})^{+}_{ij} 
& 0 & 0 & -\frac{i}{2}q_{1c} \sbar_{0i} \\
0 & -\frac{1}{2} \cD^{\ast}_{i} & 0 &\frac{1}{2}({\cal PD})^{+}_{ij} 
& -\frac{i}{2}\qt_{1c} \sbar_{0i} & 0 \\
-\frac{1}{2}\qt_{1c} \sbar_{0} & 0 & -\frac{1}{2}\qt_{1c} \sbar_{j} 
&0&0& i\Ds^{\ast} \\
0 &  \frac{1}{2} \s_{0}q_{1c} & 0& \frac{1}{2} \s_{j}q_{1c} & -i\Ds^{\ast} &0
\end{array}
\right),
\ee
where $\DM$ is defined as $\pmu -iA_{3\m}$ 
and we defined $({\cal PD})^{+}_{ij}$ as 
\be
({\cal PD})^{+}_{ij} 
= (\cD_0 \delta_{ij} - \frac{1}{2} \ep_{0ilk} \cD_l \delta_{jk}).
\ee
Only in this Abelian part, we must not forget the $\epsilon$ terms.
As we saw in section 3, $\frac{i}{2}\dm$ of
$\eta$ rows $\la$ columns should be replaced by $\de$, and
the components of $\eta$ rows $\psib$ columns are replaced 
by order $\ep$ operators $O(\ep)$ as we find the example (26).
We introduce $\cD_{\m \epd}$ as $\DM$ shifted by $\epd$ term like (25).
Then we write down $T_{t\ep}$ which defined as the $T_t$ added  
shift terms from $\pbar f_{\ep} +\eta \zeta_{\ep}$ in (19) as
\be
\left(
\begin{array}{cccccc}
-\frac{i}{2} \cD_{0 \epd} & 0 & -\frac{i}{2} \cD_{j \epd} 
& 0 & O(\ep) & O(\ep) \\
0 & -\frac{i}{2} \cD^{\ast}_{0 \epd} & 0 
& -\frac{i}{2} \cD^{\ast}_{j \epd} & O(\ep) & O(\ep) \\
-\frac{1}{2} \cD_{i} & 0 & \frac{1}{2}({\cal PD})^{+}_{ij} 
& 0 & 0 & -\frac{i}{2}q_{1c} \sbar_{0i} \\
0 & -\frac{1}{2} \cD^{\ast}_{i} & 0 &\frac{1}{2}({\cal PD})^{+}_{ij} 
& -\frac{i}{2}\qt_{1c} \sbar_{0i} & 0 \\
-\frac{1}{2}\qt_{1c} \sbar_{0} & 0 & -\frac{1}{2}\qt_{1c} \sbar_{j} 
&0&0& i\Ds^{\ast} \\
0 &  \frac{1}{2} \s_{0}q_{1c} & 0& \frac{1}{2} \s_{j}q_{1c} & -i\Ds^{\ast} &0
\end{array}
\right).
\ee
Let us path-integrate out the transversal part. 
To carry out this Gaussian integration, 
we decide the determinant of $T_{t\ep}$ as
\be
det(T_{t\ep}) \equiv det (T_{t\ep}^{\ast} T_{t\ep})^{1/2}.
\ee
(See the reference \cite{Sc}.)
When we denote the $T_{t\ep}^{\ast} T_{t\ep}$ as
\be
T_{t\ep}^{\ast} T_{t\ep}=
\left(
\begin{array}{cccc}
T^{\ast} T_{\mn}^{++} & T^{\ast} T_{\mn}^{+-} 
& T^{\ast} T_{\m \qtil}^{+} & T^{\ast} T_{\m q}^{+} \\
T^{\ast} T_{\mn}^{-+} & T^{\ast} T_{\mn}^{--} 
& T^{\ast} T_{\m \qtil}^{-} & T^{\ast} T_{\m q}^{-} \\
T^{\ast} T_{\qtil \n}^{\ +} & T^{\ast} T_{\qtil \n}^{\ -} 
& T^{\ast} T_{\qtil \qtil} & T^{\ast} T_{\qtil q} \\
T^{\ast} T_{q \n}^{\ +} & T^{\ast} T_{q \n}^{\ -} & T^{\ast} T_{q \qtil} 
& T^{\ast} T_{q q} 
\end{array}
\right),
\ee
elements of $T_{t\ep}^{\ast} T_{t\ep}$ is obtained as follows.
\be
& &( \la_{\m +} \; row \; \la_{\n +} \; column ) \nonumber \\
T^{\ast} T_{\mn}^{++} &=& (-\frac{1}{4} \cD^{2}\delta_{\mn}-F_{3\mn}^{-})
+\frac{1}{4}\qt_{1c} \s_{\m} \s_{\n} q_{1c}
+ O(\epd)
\nonumber \\
& &( \la_{\m +} \; row \; \la_{\n -} \; column ) \nonumber \\
T^{\ast} T_{\mn}^{+-}&=&0 \nonumber \\
& &( \la_{\m -} \; row \; \la_{\n +} \; column ) \nonumber \\
T^{\ast} T_{\mn}^{-+}&=&0 \nonumber \\
& &( \la_{\m -} \; row \; \la_{\n -} \; column ) \nonumber \\
T^{\ast} T_{\mn}^{--}
&=&(-\frac{1}{4} \cD^{\ast 2}\delta_{\mn}-F_{3\mn}^{-})
+\frac{1}{4}\qt_{1c} \s_{\m} \s_{\n} q_{1c}
+ O(\epd)
\nonumber \\
& &( \psib_{\qtil 2} \; row \; \la_{\n +} \; column ) \nonumber \\
T^{\ast} T_{\qtil \n}^{\ +}&=& O(\ep) \nonumber \\
& &( \psib_{\qtil 2} \; row \; \la_{\n -} \; column ) \nonumber \\
T^{\ast} T_{\qtil \n}^{\ -}
&=&(\frac{i}{4} \sbar_{\n \rho} q_{1c})^{+} \cD^{\rho} 
+ \frac{i}{2} \Ds \s_{\n}q_{1c}
+ O(\ep,\epd)
\nonumber \\
& &( \psib_{q 2} \; row \; \la_{\n +} \; column ) \nonumber \\
T^{\ast} T_{q \n}^{\ +}
&=&(-\frac{i}{4} \sbar_{\n \rho} \qt_{1c})^{+} \cD^{\rho} 
+ \frac{i}{2} \Ds \qt_{1c}\sbar_{\n}
+ O(\ep,\epd)
\nonumber \\
& &( \psib_{q 2} \; row \; \la_{\n -} \; column ) \nonumber \\
T^{\ast} T_{q \n}^{\ -}&=&O(\ep) \nonumber \\
& &(\la_{\m +} \; row \; \psib_{\qtil 2} \; column )\nonumber \\
T^{\ast} T_{\m \qtil}^{+}&=&O(\ep) \nonumber \\
& &(\la_{\m -} \; row \; \psib_{\qtil 2} \; column )\nonumber \\
T^{\ast} T_{\m \qtil}^{-}
&=&-\frac{i}{4} \cD^{\rho} (\qt_{1c} \sbar_{\rho \m} )^{+} 
-\frac{i}{2} \qt_{1c}\s_{\m}\Ds^{\ast}
+ O(\ep,\epd)
\nonumber \\
& &( \psib_{\qtil 2} \; row \;  \psib_{\qtil 2} \; column )\nonumber \\
T^{\ast} T_{\qtil \qtil}
&=&\frac{1}{4}(\sbar_{\mn}q_{1c})^{+}(\qt_{1c} \sbar^{\mn})^{+} +{\Ds^{\ast}}^2
+O(\ep^2)
\nonumber \\
& &( \psib_{q 2} \; row \;  \psib_{\qtil 2} \; column )\nonumber \\
T^{\ast} T_{q \qtil}&=&O(\ep^2)
\nonumber \\
& &(\la_{\m +} \; row \; \psib_{q 2} \; column )\nonumber \\
T^{\ast} T_{\m q}^{+}
&=&-\frac{i}{4} \cD^{\rho} (q_{1c} \sbar_{\rho \m} )^{+} 
-\frac{i}{2}\s_{\m}q_{1c}\Ds^{\ast}
+ O(\ep,\epd)
\nonumber \\
& &(\la_{\m -} \; row \; \psib_{q 2} \; column )\nonumber \\
T^{\ast} T_{\m q}^{-}&=&O(\ep) \nonumber \\
& &( \psib_{\qtil 2} \; row \;  \psib_{q 2} \; column )\nonumber \\
 T^{\ast} T_{\qtil q}&=&O(\ep^2)
\nonumber \\
& &( \psib_{q 2} \; row \;  \psib_{q 2} \; column )\nonumber \\
 T^{\ast} T_{q q}
&=&\frac{1}{4}(\sbar_{\mn}\qt_{1c})^{+}(q_{1c} \sbar^{\mn})^{+} +{\Ds^{\ast}}^2
+O(\ep^2).
\ee
When we integrate $\p$ and $\pbar$ we have to pay attention to
the shift from $\ep$ terms.
We introduce $\cD_{\m \ep}$ which include $\DM$ and shifts from $\ep$ terms
, like (24).  
After these transverse fields path-integration of (48) 
we can write this part with matrices $M$ and $T_{t \ep}$ as,
\be 
\sum_{reducible \;\; A_{\m}} \!\!\!\!\!{\cal N}det(M)^{-1/2} det(T_{t\ep}) 
det(\frac{1}{2} \cD_{\m \ep}^{\ast} \cD^{\m \ast}_{\ep} )
det(\frac{1}{2} \cD_{\m \ep}  \cD^{\m}_{\ep} )
 \langle \exp (\tilde{v} + \tau \tilde{u}) {\rangle}_{A}.
\ee
Where $\langle O {\rangle}_{A}$ means vacuum expectation value of $O$ with
Abelian Seiberg-Witten theory whose action is composed by $S_c$
and Caltan part from $\pbar f_{\ep} +\eta \zeta_{\ep}$ i.e.
\be
S_c+ \intg (\pbar f_{\ep} +\eta \zeta_{\ep})_c.
\ee
Where we denote the Caltan part of 
$\pbar f_{\ep} +\eta \zeta_{\ep}$ as $(\pbar f_{\ep} +\eta \zeta_{\ep})_c$.

If we put $\pbar f_{\ep} +\eta \zeta_{\ep}$ as an example (26), this Caltan part is
\be
(\pbar f_{\ep} +\eta \zeta_{\ep})_c &=& \epd \pbar_3 \pmu \p_3 m^{\mu}
+\epd \eta_3 \la_{3\mu} m^{\mu} -\frac{1}{2}\ep n \qt_1 \eta_3 \psib_{\qtil 1}
\nonumber \\
& & -\frac{1}{2}\ep n \psib_{q 1}\eta_3 q_1 + 
\frac{1}{2}\ep n \qt_1 \pbar_3 \p_3 q_1 -
i \ep n \psib_{q 1} \pbar_3 \psib_{\qtil 1}
\nonumber \\
&=&
\epd \brs_c (\pbar_3 \la_{3\mu} m^{\mu} ) +
 \ep \frac{i}{2} \brs_c (n\qt_1 \pbar_3 \psib_{\qtil 1} -n\psib_{q 1} \pbar_3 q).
\ee
This is the same case of Eqs.(35)(36). We find 
$\sum_{A_{\m}} \!\!
 \langle \exp (\tilde{v} + \tau \tilde{u}) {\rangle}_{A}$ is independent from $k$.
In (57), $\sum_{A_{\m}}$ is sum with weight 
${\cal N}det(M)^{-1/2} det(T_{t\ep}) 
\left| det(\frac{1}{2} \cD_{\m \ep}  \cD^{\m}_{\ep} )\right|^2$.

Let us consider the case that there is only one solution of $A_{\m}$ and $q_1$, 
there is not sum $\sum_{A_{\m}}$ and we find 
$ \langle \exp (\tilde{v} + \tau \tilde{u}) {\rangle}_{A}$ is independent from $k$.
Therefore only $det(T_{t\ep}) 
det(\frac{1}{2} \cD_{\m \ep}^{\ast} \cD^{\m \ast}_{\ep} )
det(\frac{1}{2} \cD_{\m \ep}  \cD^{\m}_{\ep} )$ term in (57) depend on $k$.
As we saw in section 3, the coefficients of $k$ have to be zero.
Then we get the non-trivial result, 
\be
\langle \exp (\tilde{v} + \tau \tilde{u}) {\rangle}_{A} =0.
\ee
But if there are several solutions of $A_{\m}$ and $q_1$,
it is unclear whether Eq.(60) is correct or not.

{ \bf non-Abelian Seiberg-Witten part}

Finally we denote non-Abelian part as
\be
{\cal N}\langle \exp (\tilde{v} + \tau \tilde{u}) {\rangle}_{nA}.
\ee
Where the connections of fixed point are restricted within irreducible
connections.
(61) is a pure non-Abelian Seiberg-Witten invariants.\\

Now the vacuum expectation value is separately written as
\be
\langle \exp (\tilde{v} + \tau \tilde{u}) {\rangle} 
&=& {\cal N}det(-4 \pi ) 
\langle \exp (\tilde{v} + \tau \tilde{u}) {\rangle}_{D} + 
{\cal N}\langle \exp (\tilde{v} + \tau \tilde{u}) {\rangle}_{nA}
 \\
+ {\lim_{\ep,\epd \rightarrow 0}}& &\!\!\!\!\!\! \!\!\!\!\!\!\!\!\!\!\!\!
\sum_{reducible A_{\m}} \!\!\!\!\!
{\cal N}det(M)^{-1/2} det(T_{t\ep}) 
det(\frac{1}{2} \cD_{\m \ep}^{\ast} \cD^{\m \ast}_{\ep} )
det(\frac{1}{2} \cD_{\m \ep}  \cD^{\m}_{\ep} )
 \langle \exp (\tilde{v} + \tau \tilde{u}) {\rangle}_{A} \nonumber 
\ee
As we saw in section 3 that this vacuum expectation value is invariant under 
changing the ratio of $\ep$ and $\epd$,
so we found that the Abelian part vanishes without 
$\left< O \right>_{R,0}= \left< O \right>_{R} \left. \right|_{\epd=0}$ in Eq.(60).
From this fact, we find following formulas,
\be
\langle \exp (\tilde{v} + \tau \tilde{u}) {\rangle}
=& & {\cal N}det(-4 \pi ) 
\langle \exp (\tilde{v} + \tau \tilde{u}) {\rangle}_{D} 
+
{\cal N}\langle \exp (\tilde{v} + \tau \tilde{u}) {\rangle}_{nA}  \\
+\left[ {\lim_{\ep \rightarrow 0}}\!\!\!\!\!\! 
\sum_{reducible A_{\m}} \!\!\!\!\!\right. & & \!\!\!\!\!\left.\left.
\frac{{\cal N}}{det(M)^{1/2}} det(T_{t\ep}) 
\left| det(\frac{1}{2} \cD_{\m \ep}  \cD^{\m}_{\ep} )\right|^2
 \langle \exp (\tilde{v} + \tau \tilde{u}) {\rangle}_{A}
 \right] \right|_{\epd=0}.\nonumber
\ee
From Eq.(37), identities are obtained as
\be
\left( \frac{\delta}{\delta k} \right)^n
\left[ 
{\lim_{\ep \rightarrow 0}}\!\!\!\!\!\! \!
\sum_{reducible A_{\m}} \!\!\!\!\!\right.\left.\left.
\frac{{\cal N}}{det(M)^{1/2}} det(T_{t\ep}) 
\left|det(\frac{1}{2} \cD_{\m \ep}  \cD^{\m}_{\ep} )\right|^2
 \langle \exp (\tilde{v} + \tau \tilde{u}) {\rangle}_{A}
\right] \right|_{k=0} =0, 
\ee
where $n=1,...,dim(ker d_A)$.
These formulas are non-trivial. These identities of U(1) topological invariants 
are obtained from SU(2) Topological QCD.
Note that vacuum expectation value of $\p$ was changed as (28),
 but detail character 
 of the shift terms did not need to get above formulas.
We comment on the Eq.(60) little more. 
This formula may imply that Abelian Seiberg-Witten invariants vanish 
in general. Indeed it is possible to apply our methods for massive topological QCD
with no obstacle. But there are some problems to identify the Abelian part 
and usual Seiberg-Witten invariants, for example Eq.(28) and there are problems
to extend to the case which has plural monopole solutions.
This subjects are discussed in \cite{sako2}.


\section{Conclusions}
We have studied massless topological QCD 
in detail and found new relations (63) and (64).
One of the results of excluding mass terms is that our theory does not have
spontaneous gauge symmetry breaking phase in usual meaning.
Hence, we could not distinguish between 
reducible and irreducible connections without no modification.
We gazed 
$det(D^{\m}\dm) =0$ when connections are reducible.
Infinitesimal shift terms are introduced to the Lagrangian to account 
zero-eigenvalue states of $D^{\m}\dm$.
For this terms we could contain the Abelian part and treat 
separately reducible and irreducible connections. 
That determinant are obtained from $\p$ and $\pbar$ integral. 
Fermionic integral of $\eta$, which is super partner of $\pbar$, 
$\la$, $\x$ etc. cause the determinant of $T$.  
This determinant offset $det(D^{\m}\dm)=0$ .
Infinitesimal shift terms is added to the Lagrangian 
to shift the both zero-determinants.
We could change the ratio of infinitesimal shift terms 
without 
changing vacuum expectation value.
As a result of this, we got formulas (63) and (64) (and especially case Eq.(60)).
These are non-trivial relations between topological invariants.
The identities of U(1) topological invariants are obtained 
from BRS symmetry of Topological QCD like Ward-Takahashi identity.

When we interpret Topological Field Theory as a gauge fixing theory like
\cite{B.M}, the zero-modes of $\dm$ is interpreted as Gribov zero-modes.
As we saw in section 4, Gribov zero-modes break BRS symmetry often and 
topological symmetry breaking occur. So the external fields in the section 4
avoid topological symmetry breaking from Gribov zero-modes.

The next subjects we have to investigate are to caluculate actually in 
some models and to ascertain these formulas.
We studied only SU(2) gauge theries in the present paper, so we want to extend 
it more general case.
To carry it out in our formalism, we have to construct
the tool that embed the equations to classify the reducible connections
in equations of motion from some topological action.
This is one of our future problems.\\
The relation between the 
$\langle \exp (\tilde{v} + \tau \tilde{u}) {\rangle}_{A}$ 
and usual Seiberg-Witten invariants have to be studied more carefully.
$\langle \exp (\tilde{v} + \tau \tilde{u}) {\rangle}_{A}$ is topological 
invariants but have some difference from usual Seiberg-Witten invariants.
It is important problem to make the difference clear.\\

{ \Large \bf Acknowledge.}\\
 I am grateful to Professor K.Ishikawa for 
helpful suggestions and observations and a critical reading of the manuscript.


\appendix
\section{  \ \ \ \  Reducible connection}

In this appendix, we summarize some basic of reducible connections
for physicists who is unfamiliar with this words.
For the convenience, we treat the only case of SU(2) gauge group.
We consider a connection on a point $p$ on a back ground manifold $M$.
A holonomy group is defined as subgroup of SU(2) which 
transform the connection as parallel trnsformations around any loop
with the start point $p$. Therefore, holonomy groups
are understood as groups which is needed actually to introduce 
each connection.
We put $H$ as the holonomy group.
We can define the reducible connection with the holonomy group which is 
classified in following two cases.

(1) $ H\subseteq \left\{ \pm 1\right\} $.

(2) $ H $ is conjugate to U(1) .

In the first case connections are flat and this case
is realized when the Chern number is zero.
In our theory, case (1) is ignored.
When connections do not satisfy above each conditions then we
call them irreducible connections, and centralizer of $H$ is $\{ \pm 1 \}$.

We can understand the relation,
which is used many times in this paper,
 between reducibility and zero-modes of $d_A : 
Ad \eta \bigotimes {\bigwedge}_{0} \rightarrow 
 ad \eta \bigotimes {\bigwedge}_{1}$
as follows. 
If $d_A$ has a zero-mode $\p_0$, we obtain an one parameter group 
$\{ \exp(t \p_0 ) \; | \;t \in {\bf R} \; d_A \p_0 =0 \}$ 
whose elements transform the connection identically because
\be
g^{-1}dg +g^{-1}A g=td_A \p_0 =0 .
\ee
This means that centralizer of $H$ is not $\{ \pm 1 \}$ and 
this connection is reducible.
Oppositely if a connection is reducible, then there is an one parameter group
whose elements satisfy
\be
 g^{-1}dg +g^{-1}A g=0 
\ee
 and $g \neq \pm 1$.
We obtain 
\be
d_A \left( \frac{d g}{dt} \right) =0 
\ee
after differentiate (60) by the parameter $t$.
Therefore we understand $\frac{d g}{dt}$ is a zero mode of $d_A$.
 
Next we will see the process of reducible connection defined as 
\be
A=\left( \begin{array}{cc}
A_L & 0 \\
0 & A^{\ast}_L
\end{array} \right)
\ee
where $A_L$ is a connection on a complex line bundle $L$ and
$A^{\ast}_L$ is a its complex conjugate.
We can put the orthogonal basis of 2-dimensional complex 
vector bundle, 
\be
q= \left(
{\begin{array}{c}
q_{1} \\
q_{2}
\end{array}} \right),
\ee
as 
\be
e_1= \left(
{\begin{array}{c}
q_{1} \\
0
\end{array}} \right), \ \ \ 
e_2= \left(
{\begin{array}{c}
0 \\
q_{2} 
\end{array}} \right). 
\ee
We introduce complex line bundle $L$ and $L'$ 
with a parallel trnsformation operator
$P_l$ where $l$ is represented as some loop as 
\be
L= \{ c P_l(e_1) | c \in {\bf C} \} \\
L'= \{ c P_l(e_2) | c \in {\bf C} \} .
\ee
In our theory, reducible connections means U(1) connections and 
we can take $P_l$ as diagonal matrices.
So we find that the definitions of $L$ and $L'$ are 
unrelated of loop $l$.
We could established $L$ and $L'$ in this way.
Therefore connection $A$ can be represented as
\be
A=A_L \oplus A_{L'}
\ee
where $A_L$ and $A_{L'}$ are connections of $L$ and $L'$ respectively.
Since $A$ have to be valued in u(1) now, $A$ is pure imaginary 
i.e. $A_{L'}=A_{L}^{\ast}=-A_{L}$.
We obtain (61) and 
\be
F_A=\left( \begin{array}{cc}
dA_L & 0 \\
0 & -dA_L
\end{array} \right).
\ee
In this paper, we used these results in section 4.


\section{  \ \ \ \  Gaussian Integral}

In the section 4, 
we integrate fermionic fields in the Abelian Seiberg-Witten part.
The methods of integration used there are studied in reference
\cite{P.R.}\cite{Sc}.
It is possible to do this integration in not only 
Abelian case but also other case.
For the non-Abelian parts, we can do it more general.
i.e. we  put classical back ground fields 
as $A_{c \m}^{a} T_a$ and $q_c$ in generally
and expand $S_{QCD}$ to second order of quantum fields. 
We write down the result of fermionic part as follows.
\be 
(\eta_{a} \x_{0i a} \psi_{q}^{t} \psi_{\qtil})
 \left( {\large T} \right)
\left( \begin{array}{c} \la_{0 b}  \\
\la_{j b} \nonumber \\
\psib_{\qtil} \nonumber \\
\psib_{q}^{t} \end{array} \right),
\ee
and $T$ is concretely 
\be
\left(
\begin{array}{cccc}
-iTr (T_a D_0 T_b) & -iTr (T_a D_j T_b)  & 0 & 0 \\
-Tr(T_a D_i T_b) & Tr(T_a(PD)^{+}_{ij}T_b)&i\qt_{c} 
\sbar_{0i} T_a & -iq_c^t \sbar_{0i} T_a^t \nonumber \\
-T_a^t \sbar_{0} (\qt_{c})^t & -T_b^t \sbar_{j} (\qt_{c})^t 
&0&i\not \!\!D \nonumber \\
\s_{0} T_b q_{c} & \s_{j} T_b q_{c} & -i\not \!\!D  &0
\end{array}
\right).
\ee
Where we denote $(D_0 \delta_{ij} - \frac{1}{2} \ep_{0ilk} D_l \delta_{jk})$ 
as $(PD)^{+}_{ij}$.
We chose space elements of self dual field $\x$ as substantial elements.
When we take acoount of $\ep$ shift terms, then the first row
of $T$ change in order $\ep$.
We call this shifted $T$ as $T_{\ep}$.
For example, we obtained $T_{\ep}$ of the case concretely as
\be
\left(
\begin{array}{cccc}
-iTr (T_a (D_0 -\epd m_0 ) T_b) & -iTr (T_a (D_j -\epd m_j)T_b)  
& -\ep n \qt_c T_a & -\ep n q^t_c T^t_a \\
-Tr(T_a D_i T_b) & Tr(T_a(PD)^{+}_{ij}T_b)&i\qt_{c} \sbar_{0i} T_a 
& -iq_c^t \sbar_{0i} T_a^t \nonumber \\
-T_a^t \sbar_{0} (\qt_{c})^t & -T_b^t \sbar_{j} (\qt_{c})^t &0
&i\not \!\!D \nonumber \\
\s_{0} T_b q_{c} & \s_{j} T_b q_{c} & -i\not \!\!D  &0
\end{array}
\right).
\ee

$T$ is not a map from a space into itself.
So we have to pay attention to define its determinant.
We put adjoint operator of $T$ as $T^{\ast}$, and
define $det(T)$ as
\be
det(T) \equiv det (T^{\ast} T)^{1/2}.
\ee
(See the references \cite{P.R.} \cite{Sc}.)
Now we can treat $(T^{\ast} T)$ with not space indices $i$ but
space-time indices $\m$.
We name the elements of $(T^{\ast} T)$ by
\be
&(&\la_{\m}^{c} \; \psib_{\qtil} \; \psib_{q} )
 \left( {\large T^{\ast}T }\right)
\left( \begin{array}{c} \la_{\n}^{b} \nonumber \\
\psib_{\qtil} \nonumber \\
\psib_{q} \end{array} \right) ,\\
T^{\ast} T&=&
\left(
\begin{array}{ccc}
T^{\ast} T_{\mn}^{cb} & T^{\ast} T_{\m \qtil}^{c} & T^{\ast} T_{\m q}^{c} \\
T^{\ast} T_{\qtil \n}^{\ b} & T^{\ast} T_{\qtil \qtil} 
& T^{\ast} T_{\qtil q} \\
T^{\ast} T_{q \n}^{\ b} & T^{\ast} T_{q \qtil} & T^{\ast} T_{q q} 
\end{array}
\right)
\ee
The elements of $(T^{\ast} T)$ is obtained as follows.
\be
& &( \la_{\m}^{c} \; row \; \la_{\n}^{b} \; column ) \nonumber \\
T^{\ast} T_{\mn}^{cb} &=&-\frac{1}{2} 
Tr \; (T_c (D^{2}\delta_{\mn}-F_{\mn}^{-})T_b)
+\qt_{c}T_b T_c \s_{\m} \s_{\n} q_{c} 
+\qt_{c}T_c T_b \s_{\m} \s_{\n} q_{c} 
\nonumber \\
& &( \psib_{\qtil} \; row \; \la_{\n}^b \; column ) \nonumber \\
T^{\ast} T_{\qtil \n}^{\ b}&=&\frac{i}{2} 
(T_a \sbar_{\n \rho} q_{c})^{+} (2Tr \; T_a  D^{\rho} T_b)
 +i \not \!\!D \s_{\n} T_b q_{c}
\nonumber \\
& &( \psib_{q}^t \; row \; \la_{\n}^b \; column ) \nonumber \\
T^{\ast} T_{q \n}^{\ b}&=&(-\frac{i}{2} 
\sbar_{\n \rho}T_a^t (\qt_{c})^t)^{+} D^{\rho} 
+ i\not \!\!D T_a^t \sbar_{\n} (\qt_{c})^t
\nonumber \\
& &(\la_{\m}^c \; row \; \psib_{\qtil} \; column )\nonumber \\
 T^{\ast} T_{\m \qtil}^{c}
&=&i(Tr \;T_c D^{\rho} T_a) (\qt_{c} \sbar_{\rho \m}T_a )^{+} 
-i\qt_{c}T_c\s_{\m}\not \!\! D
\nonumber \\
& &( \psib_{\qtil} \; row \;  \psib_{\qtil} \; column )\nonumber \\
T^{\ast} T_{\qtil \qtil}&=&(T_a \sbar_{\mn}q_{c})^{+}
(\qt_{c} \sbar^{\mn} T_a)^{+} +{\not \!\!D}^2
\nonumber \\
& &( \psib_{q} \; row \;  \psi_{\qtil} \; column )\nonumber \\
 T^{\ast} T_{q \qtil}&=&
(T_a^t \sbar_{\mn}(\qt_{c})^t)^{+}(q_{c}^t \sbar^{\mn} T_a^t)^{+} 
\nonumber \\
& &(\la_{\m}^c \; row \; \psib_{q}^t \; column )\nonumber \\
T^{\ast} T_{\m q}^{c}
&=&-i(Tr \;T_c D^{\rho} T_a)(q^t_{c} \sbar_{\rho \m}T_a^t )^{+}
 -i(q_{c})^t T_c^t \sbar_{\m}\not \!\!D
\nonumber \\
& &( \psib_{\qtil} \; row \;  \psib_{q}^t \; column )\nonumber \\
T^{\ast} T_{\qtil q}&=&-(T_a \s_{\mn}q_{c})^{+}(q_{c}^t \s^{\mn} T_a^t)^{+} 
\nonumber \\
& &( \psib_{q}^t \; row \;  \psib_{q}^t \; column )\nonumber \\
T^{\ast} T_{q q} 
&=&(\sbar_{\mn}T_a^t (\qt_{c})^t)^{+}(q_{c}^t T_a^t \sbar^{\mn})^{+}
+{\not \!\!D}^2.
\ee
When we compares this to the matrices of the section 4, 
we understand that differences are in only covariant derivative $D_{\m}$
and $\Dm$.
Note that matrix obtained from bosonic part is 
almost same as $T^{\ast} T$.


\end{document}